\documentclass[aps,nofootinbib,twocolumn,prd,eqsecnum,showpacs,showkeys,preprintnumbers,,altaffilletter]{revtex4-1}
\usepackage[caption=false]{subfig}
\usepackage{graphicx}
\usepackage{amsmath}
\usepackage{amsfonts}
\usepackage{amssymb}
\usepackage{accents}
\usepackage{mathbbol}
\usepackage{float}
\usepackage{color}
\usepackage{bm}
\usepackage{mathrsfs}
\usepackage{epstopdf}
\usepackage{url}
\usepackage{footnote}
\usepackage{textcomp}
\usepackage{natbib}

\begin{document}


\title{Phantom Dark Ghost in Einstein-Cartan Gravity}
\author{Yu-Chiao Chang$^{1,2}$}
\email{f95222075@ntu.edu.tw}
\author{Mariam Bouhmadi-L\'{o}pez$^{3,4}$}
\email{{\mbox{mariam.bouhmadi@ehu.eus}}}
\author{Pisin Chen$^{1,2,5,6}$}
\email{pisinchen@phys.ntu.edu.tw}
\date{\today}

\affiliation{
${}^1$Department of Physics, National Taiwan University, Taipei, Taiwan 10617\\
${}^2$LeCosPA, National Taiwan University, Taipei, Taiwan 10617\\
${}^3$Department of Theoretical Physics, University of the Basque Country
UPV/EHU, P.O. Box 644, 48080 Bilbao, Spain\\
${}^4$IKERBASQUE, Basque Foundation for Science, 48011, Bilbao, Spain\\
${}^5$Graduate Institute of Astrophysics, National Taiwan University, Taipei, Taiwan 10617\\
${}^6$Kavli Institute for Particle Astrophysics and Cosmology, SLAC National Accelerator Laboratory, Stanford University, Stanford, CA 94305, U.S.A.
}

\begin{abstract}
A class of dynamical dark energy models is constructed through an extended version of fermion fields \textcolor{black}{corresponding to phantom dark ghost} spinors, which are spin one half with mass dimension one. We find that if \textcolor{black}{these spinors interact} with torsion fields in a homogeneous and isotropic universe, then it does not imply any future dark energy singularity or any abrupt event, though the fermion has a negative kinetic energy. In fact, the equation of state of this dark energy model will asymptotically approach the value $w=-1$ from above without crossing the phantom divide and inducing therefore a de Sitter state. Consequently, we expect the model to be stable because no {\color{black} \it{real}} phantom fields will be created. At late time, the torsion fields will vanish as the  \textcolor{black}{corresponding phantom dark ghost} spinors dilute. As would be expected, intuitively, this result is unaffected by the presence of cold dark matter although the proof is not as straightforward as in general relativity.
\end{abstract}



\maketitle

\section{Introduction}
It is well-known that General relativity (GR) is a successful theory in agreement with a great number of observations \cite{Will2006}. It describes gravity by means of the Einstein-Hilbert action, which is proportional to the curvature scalar $R$. When this action is varied with respect to the Riemannian metric $g_{\mu\nu}$, one obtains the Einstein equations. It is also well-known that there are two conditions assumed about the affine connection in GR, one is the metric compatibility, and the other is the torsion-free condition. Under these two conditions, there exists exactly one connection, namely the Levi-Civita connection, or sometimes the Christoffel connection, on a given manifold with a given metric. On the other hand, relativistic quantum field theory (QFT) is a highly successful theory in describing the electromagnetic, strong, and weak interactions. However, the framework of QFT is basically constructed in the flat Minkowski spacetime and interactions are independent of the background spacetime. From Einstein's GR, we know that spacetime itself should be dynamical and interact with other fields through the gravitational field. Following Einstein's point of view, the gravitational field should couple with all the fields in our universe. Therefore, angular momentum should also have a contribution to the energy-momentum tensor. Hence, spinor fields, which carry intrinsic angular momentum predicted by QFT, may actually couple with gravity. Specifically, we know that in QFT, all elementary particles are classified by means of the irreducible unitary representations of the Poincar\'{e} group which has two labels, the mass $m$ and the spin $s$. In ordinary GR, mass couples to the curvature, whereas spin does not couple to any geometrical quantity \cite{Hehl1976}. To treat mass and spin on equal footing, spin should couple to spacetime metric in some way source  {\color{black} the} gravitational field. {\color{black} This would lead to a theory of gravity more general than GR.} Einstein-Cartan theory is probably the simplest extention of GR {\color{black} which incorporate spin and mass in its formulation, providing, therefore, a more complete treatment in what referes to the Poincar\'{e} group}\footnote{{\color{black}Similar to the situation in GR where not only the mass but also the linear momentum contribute to the energy-momentum tensor, i.e., not only $\rho$ (energy density) but also $\mathbf{p}$ (linear momentum)  contribute, in the case of spins, a generalized theory of gravity that honours the Poincar\'{e} group should not only include the intrinsic spin but also the orbital angular momentum and their contributions to gravity, somewhat similar to what is done in Quantum Mechanics.}}.

Recently, Ahluwalia-Khalilova and Grumiller proposed a novel four-component spinors by means of the eigenspinors of the charge conjugation operator $C$ in momentum space \cite{Ahluwalia-KhalilovaD.V.2005, Ahluwalia-Khalilova2005, Ahluwalia2009}.
{\color{black}They can be named dark spinors \cite{Bohmer2008}.}
They satisfy $(CPT)^2=-\mathbb{1} $ and have the dimension of mass. In comparison, the Dirac spinors in the standard model satisfy $(CPT)^2= \mathbb{1}$ with the dimension of mass $3/2$. {\color{black} Since CPT is anti-unitary for the dark spinor, it restricts the interactions between it and other standard model particles, that is, the interactions between a dark spinor and the standard matter particles will always be paired in a dark spinor and its conjugate. Because the dark spinors have dimension of mass, by power counting renormalizability, interactions of the dark spinors with other particles of the standard model take place only through the Higgs doublet or with gravitational fields. Hence, these spinors are candidate {\color{black} for} dark matter \cite{Ahluwalia-KhalilovaD.V.2005, Ahluwalia-Khalilova2005}, this is in part the origin of its name. Besides, these spinors can couple to all parts of the torsion tensor \cite{Ahluwalia-KhalilovaD.V.2005, Bohmer2006, Bohmer2008} unlike the ordinary Dirac spinors that can only couple to its axial vector part \cite{Hehl1976, Bohmer2006, Popp2012, Popawski2010a}. Hence, the dark spinors may provide far more interesting implications in cosmology than ordinary spinors.} 

On the other hand, it has been shown that our universe is spatially flat and started accelerating in the recent past. This conclusion has been backed up by many observational data such as type Ia supernovae \cite{Betoule2014}, baryon acoustic oscillation \cite{Bailey}, cosmic microwave background radiation \cite{PlanckCollaboration2015}. To explain the recent accelerated expansion, an unknown component with negative pressure is usually assumed, and called dark energy. In the last years, answering what dark energy really is has become one of the most challenging questions in cosmology. The simplest candidate for dark energy is a small positive cosmological constant $\Lambda$ which gives the equation of state $w\equiv p/\rho=-1$ where $p$ stands for pressure and $\rho$ for the dark energy density. Although the cosmological constant with cold dark matter; i.e. $\Lambda$CDM model, can explain pretty well the observational data, it suffers, however, from fine-tuning and coincidence problems, in other words, why the cosmological constant is so small and only became dominating almost at present? To address these issues, cosmologists have considered dynamical dark energy models, such as quintessence \cite{Ratra1988}, phantom \cite{Caldwell2002}, quintom \cite{Feng2005}. In these models, the equation of state $w$ is not necessarily a constant and may evolve with time. Most dark energy models are constructed by scalar fields, having $w\geq-1$, converging to $w=-1$, and the quantum stability of such theories is guaranteed by the energy conditions \cite{MattVisser}. However, recent models with equation of state $w<-1$ and converging to $w=-1$ from below, generally referred to as phantom, have drawn lots of attention. The equation of state $w<-1$ is usually realized by a negative kinetic energy, and this counter intuitive assumption violates all the energy conditions, resulting usually in 
singularities {\color{black} or abrupt events}, such as the big rip \cite{Starobinsky2000,Caldwell2003,Caldwell2002,Gonzalez-Diaz2004,Gonzalez-Diaz2004a}, the sudden \cite{Gorini2004,Barrow2008,Nojiri2005,Bouhmadi-LopezMariamPedroF.Gonzalez-Daz2008} the big freeze \cite{Nojiri2005,Bouhmadi-LopezMariamPedroF.Gonzalez-Daz2008,Nojiri2005a,Odintsov2004}, the type-IV singularity \cite{Nojiri2005,MariamBouhmadi-LopezPedroF.Gonzalez-Diaz2008,Nojiri2005a,Odintsov2004,Nojiri2008,Bamba2008}, the little rip \cite{Ruzmaikina1970,Stefancic2005,Bouhmadi-Lopez2008,Frampton2011}, or the little sibling of the big rip \cite{Bouhmadi-Lopez2014a}. Nevertheless, a phantom model with equation of state $w<-1$ could still be a phenomenologically viable model, for example, as an effective description \cite{Carroll2003,Kahya2007}.

Dark spinor fields as a source of dark energy have been considered by different authors \cite{Bohmer2008,Bohmer2008a,Bohmer2011,Christian2009,Basak2012,Fabbri2012,Bohmer2010,WeiHao2011}. However, none of them consider the spinor field with \textcolor{black}{a} negative kinetic energy and torsion at the same time. Our work addresses the following question: Can a phantom \textcolor{black}{dark ghost} spinor field embedded in Einstein-Cartan gravity avoid dark energy singularities? Technically, this amounts to asking what would be the effect of torsion in phantom dark energy models. We find that in this case the equation of state will not cross the phantom divide. In other words, we provide an example of a dark energy model with a negative kinetic energy without any potential to produce quantum instabilities as it fulfils the null energy condition.

This paper is organized as follows. In Section II, we review the basic ideas of Einstein-Cartan gravity.
In Section III, we construct a dark energy model \textcolor{black}{based on a dark ghost} spinor interacting with torsion in a homogeneous and isotropic universe. In Section IV, we compute the field equations of this model and qualitatively solve them, studying the evolution of the Hubble parameter $H=\frac{\dot{a}}{a}$ and the equation of state $w_{de}$ for the \textcolor{black}{dark ghost} spinors. Furthermore, we consider the existence of another component corresponding to cold dark matter (CDM). We solve  the two components system numerically. Then, we study the behavior of the equations of state for dark energy, dark matter, the torsion fields as well as the effective equation of states whenever necessary. In section V, we conclude and discuss our results.

\section{Summary of Einstein-Cartan Theory}
Einstein-Cartan(-Sciama-Kibble) theory of gravity, like GR, is based on  Einstein-Hilbert action \cite{Hehl1976}. It relaxes, however, the GR constraint on the affine connection, $\tilde{\Gamma}^{k}_{ij}$, to be symmetric in its lower two indices. Hence the anti-symmetric part of the affine connection, i.e. the Cartan torsion tensor
${S_{ij}}^k=\tilde{\Gamma}_{[ij]}^k=\frac{1}{2}(\tilde{\Gamma}_{ij}^k-\tilde{\Gamma}_{ji}^k)$, which is a dynamical variable, independent of the Riemannian metric $g_{ij}$ is also allowed \cite{Hehl1976}. The notation $[ij]$ stands for the anti-symmetrization of the tensor indices, defined by $T_{[ij]}=\frac{1}{2}(T_{ij}-T_{ji})$ for any tensor $T_{ij}$; similarly, the notation $(ij)$  means symmetrization of the tensor indices, $T_{(ij)}=\frac{1}{2}(T_{ij}+T_{ji})$. Quantities denoted with a tilde always take torsion into account. The torsion tensor has 24 independent components in general. Note that we still require the metric compactibility condition $\tilde{\nabla}^\rho g_{\mu\nu}=0$, and the metric compactible affine connection with torsion can be written as \cite{Hehl1976}
\begin{equation}
\tilde{\Gamma}_{ij}^k=\Gamma_{ij}^k-{K_{ij}}^k,
\end{equation}
where $\Gamma_{ij}^k$ is the usual Christoffel symbol, defined by $\Gamma_{ij}^k=\frac{1}{2}g^{kl}(\partial_ig_{lj}+\partial_jg_{il}-\partial_lg_{ij})$,
and ${K_{ij}}^k$ is called the contortion tensor, defined by \cite{Hehl1976}
\begin{equation}
{K_{ij}}^k=-S_{ij}{}{}^k-2S^k{}_{(ij)}=-S_{ij}{}{}^k-S^k{}_{ij}-S^k{}_{ji}.
\end{equation}
Note that the Cartan torsion tensor is anti-symmetric in its first two indices, ${S_{ij}}^k=-{S_{ji}}^k$, by definition; however, the contortion tensor is anti-symmetric in its last two indices, ${K_{ij}}^k=-{{K_i}^k}_j$. By virtue of the last two equations, the inverse relation between the torsion and the contortion tensor reads ${S_{ij}}^k=-{K_{[ij]}}^k$.

After introducing the Cartan torsion and the contortion tensor, we can now define the action of the Einstein-Cartan theory of gravity which is simply the Einstein-Hilbert action with torsion and metric which are regarded as independent variables:
\begin{equation}
S=\int d^4x \sqrt{-g}\left( \frac{1}{2\kappa}\tilde{R}+\tilde{\mathcal{L}}_m \right),
\end{equation}
where we set the speed of light to be unity, $c=1$, the gravitational coupling constant $\kappa={8\pi G}$, and $\tilde{\mathcal{L}}_m$ is the lagrangian density of matter minimally coupled to gravity. Before taking the variation of the action, it should be noted that the independent variables are the metric tensor $g_{ij}$ and the torsion tensor ${S_{ij}}^k$, the contortion tensor ${K_{ij}}^k$ actually depends on the metric since we lower and rise some indices via $g_{ij}$ \cite{Hehl1976}. Even though, in principle we should do the variation with respect to the metric and the torsion tensors, it is more convenient to vary with respect to the contortion tensors instead, since the affine connection can be separated into the torsion-free Christoffel symbol and the contortion tensor, and the relation between torsion and contortion is only algebraic. Thus, we will vary the total action with respect to the metric and the contortion tensors, and we obtain two field equations:
\begin{gather}
\tilde{R}_{ij}-\frac{1}{2}\tilde{R}g_{ij}=\kappa \tilde{\Sigma}_{ij}, \\
{S^{ij}}_k+\delta^i_k{S^j}_l{}^l-\delta^{j}_{k}{S^i}_l{}^l=\kappa {\tau^{ij}}_k,
\end{gather}
where the first field equation is similar to the original Einstein equation, we define $\tilde{G}_{ij}\equiv \tilde{R}_{ij}-\frac{1}{2}\tilde{R}g_{ij}$, which is the Einstein tensor with torsion, $\tilde{\Sigma}_{ij}$ is the canonical energy-momentum tensor,  and the second one is called the Cartan equation. Note that in general, $\tilde{R}_{ij}$ is no longer symmetric, so as the $\tilde{G}_{ij}$ due to the fact that affine connection is asymmetric $\tilde{\Gamma}_{ij}^k\neq \tilde{\Gamma}_{ji}^k$. We define the modified torsion tensor to be $T^{ij}{}{}_k \equiv {S^{ij}}_k+\delta^i_k{S^j}_l{}^l-\delta^{j}_{k}{S^i}_l{}^l $. The right hand side (rhs) of Eq.(2.5) is the spin tensor $\tau^{ij}{}{}_k$, which is defined by
\begin{equation}
\tau_k{}{}^{ji}=\frac{\delta\tilde{\mathcal{L}}_m}{\delta K_{ij}{}{}^k}.
\end{equation}
The canonical energy-momentum tensor is given by
\begin{equation}
\tilde{\Sigma}_{ij}=\tilde{\sigma}_{ij}+ \left( \widetilde{\nabla} + K_{lk}{}{}^l \right) \left( \tau_{ij}{}{}^k-\tau_j{}^k{}_i+\tau^k{}{}_{ij} \right),
\end{equation}
where $\tilde{\sigma}_{ij}$ is the metric energy-momentum tensor, defined by
\begin{equation}
\tilde{\sigma}_{ij}=\frac{2}{\sqrt{-g}}\frac{\delta \left( \sqrt{-g}\, \tilde{\mathcal{L}}_m \right)}{\delta g^{ij}},
\end{equation}
and the second term in Eq. (2.7) is the correction to the energy-momentum tensor generated by {\color{black} the} spin-torsion interaction.
Since the Cartan equation is, in general, a set of 24 algebraic equations rather than differential relations between torsion and spin fields, it means that there would be no torsion outside matter distribution with spin source. In other words, torsion cannot propagate through the spacetime outside the matter distribution with spin source \cite{Hehl1976}.
Furthermore, we are able to substitute the torsion everywhere by the spin {\color{black} tensor} and eliminate the torsion from the formalism. It then leads to the so-called Einstein-Cartan equation,
\begin{equation}
G_{ij}=\kappa \widehat{\sigma}_{ij},
\end{equation}
where the effective energy-momentum tensor on rhs is given by \cite{Hehl1976,Popawski2014}
\begin{align*}
\widehat{\sigma}_{ij} &\equiv \tilde{\sigma}_{ij}+ \kappa \left( -4\tau_i{}^k{}_{[ l} \tau_{|j|}{}^l{}_{k ]}-2\tau_i{}^{kl}\tau_{jkl}+\tau^{kl}{}{}_i\tau_{klj}\right)\nonumber\\
&+\frac{1}{2}\kappa g_{ij}\left(4\tau_m{}^k{}_{[l} \tau^{ml}{}{}_{k]}+\tau^{klm}\tau_{klm} \right)\nonumber\\
&\equiv \tilde{\sigma}_{ij}+\kappa u_{ij}, \tag{2.10}
\end{align*}
which is symmetric and obeys the usual conservation law $\nabla^j \hat{\sigma}_{ij}=0$.
In fact, note that the Einstein-Cartan equation can be rewritten without including any torsion term by simply substituting all the torsion terms with the spin tensor terms. For example, Eq. (2.10) can be rewritten without any torsion term as the metric energy-momentum tensor can be split as a pure metric term plus a spin tensor term.
One can interpret Eqs. (2.9) and (2.10) as that the geometry is a result from the contribution of the matter field plus some spin-spin interaction. In summary, all the torsion terms disappear in both side of Eq. (2.9), however, torsion exists on both sides of Eq. (2.4).

\section{A Dark Energy Model Of Phantom Dark Ghost Spinors With Torsion}
In this section, we consider a dynamical dark energy model constructed from {\color{black} dark spinors in Einstein-Cartan theory. In fact,  we will consider dark ghost spinors (cf. the action (3.8)).} To begin with, since it is sometimes more convenient to work in an orthonormal basis, let us introduce the vielbein $e^\mu_a$, defined by
\begin{equation}
g_{\mu\nu}e^\mu_ae^\nu_b=\eta_{ab},
\end{equation}
where $g_{\mu\nu}$ is the spacetime metric and $\eta_{ab}$ is the metric of the local inertial frame given by $\eta_{ab}=\textrm{diag}(1,-1,-1,-1)$. The Greek letters $(\mu, \nu,\ldots)$ take values $(t, x,\ldots)$ and are called the \textit{holonomic} indices representing the spacetime frame, the Latin letters $(a, b,\ldots)$ taking values $(0,1,\ldots)$ are called the \textit{anholonomic} indices representing the local inertial (orthonormal) frame. We choose the anholonomic $\gamma$-matrices, $\gamma^a$, in the Weyl representation \cite{Bohmer2006}
\begin{equation}
\gamma^0=\left(
           \begin{array}{cc}
             \mathbb{O} & \mathbb{1} \\
             \mathbb{1} & \mathbb{O} \\
           \end{array}
         \right), \,\,\,
\gamma^i=\left(
           \begin{array}{cc}
             \mathbb{O} & -\sigma^i \\
             \sigma^i  & \mathbb{O} \\
           \end{array}
         \right), \,\,\,
\gamma^5=\left(
           \begin{array}{cc}
             \mathbb{1} & \mathbb{0} \\
             \mathbb{0} & \mathbb{-1} \\
           \end{array}
         \right),
\end{equation}
where $i=1,2,3$, {\color{black} $\sigma^i$ are the Pauli matrices} and $\gamma^5=i\gamma^0\gamma^1\gamma^2\gamma^3$. The $\gamma$-matrices satisfy
\begin{equation}
\{\gamma^a,\gamma^b\}=2\eta^{ab}.
\end{equation}
We define $\gamma^\mu=e^\mu_a\gamma^a$, then $\{\gamma^\mu,\gamma^\nu\}=2g^{\mu\nu}$. The anti-commutator of two matrices is defined as: $\{A,B\}=AB+BA$ while the commutator as $[A,B]=AB-BA$.

The covariant derivatives of the \textcolor{black}{dark ghost} spinor $\lambda$ and its dual $\accentset{\neg}{\lambda}$ in the local inertial frame are defined in the same way as for the ordinary spinors, i.e.
\begin{align}
\nabla_\mu \lambda &= \partial_\mu \lambda - \Gamma_\mu \lambda, \tag{3.4a} \\
\nabla_\mu \accentset{\neg}{\lambda} &= \partial_\mu \accentset{\neg}{\lambda} + \accentset{\neg}{\lambda}\Gamma_\mu, \tag{3.4b}
\end{align}
where $\Gamma_\mu$ is called the spin connection which is used to make the covariant derivative of a spinor transform correctly under both local Lorentz transformation and general coordinate transformation. \textcolor{black}{In addition,  the dual of the dark ghost spinor is defined as}
\begin{equation}
\accentset{\neg}{\lambda}_\alpha (\mathbf{p})= i \varepsilon^\beta_\alpha \lambda_\beta^{\dag}(\mathbf{p})\gamma^0, \tag{3.5}
\end{equation}
 \textcolor{black}{with the antisymmetric symbol $\varepsilon^{\{-,+\}}_{\{+,-\}}=-1=-\varepsilon^{\{+,-\}}_{\{-,+\}}.$}
\textcolor{black}{It should be mentioned that  this definition of dual has been recently replaced by D. Ahluwalia in Ref. \cite{Ahluwalia2016aa,Ahluwalia2016bb} in order to remove problems related to  Lorentz
violation and locality concerning Eq. (3.5). In addition, in what refers to cosmological applications, at a classical level, the use of Eq. (3.5) is completely fine.}
Note that since the {\color{black}dark} spinor is still of the form $(1/2,0)\oplus(0,1/2)$ within the representation of the Lorentz group,
{\color{black} it is consequently} a spin 1/2 particle and not a spin 3/2 particle. So, the covariant derivative in {\color{black}Eq. (3.4)}  is
covariant. By further requiring that $\nabla_\mu e_\nu ^a =0$, the relation between the spin connection and the affine connection can be obtained in the following form \cite{Bohmer2006,Bohmer2010}
\setcounter{equation}{7}
\begin{gather}
\Gamma_\mu=\frac{i}{2}\omega_{\mu}^{ab}f_{ab}, \tag{3.6a}\\
\omega_{\mu}^{ab}= e^a_\nu\partial_\mu e^{\nu b}+e^a_\nu e^{\sigma b}\Gamma_{\mu\sigma}^\nu, \tag{3.6b}
\end{gather}
where $f^{ab}=\frac{i}{4}[\gamma^a,\gamma^b]$ is the generator of the local Lorentz group. Within the presence of torsion fields, we now need to extend the definition of the covariant derivatives on spinors to include torsions. According to Eq. (2.1), we may separate the non-torsion free affine connection into a torsion-free Christoffel symbol plus a contortion tensor. Applying this relation into the spin connection Eq. (4.4) and after some algebra, we obtain:
\setcounter{equation}{5}
\begin{equation}
\tilde{\nabla}_a \lambda= \nabla_a \lambda + \frac{1}{4}K_{abc}\gamma^b\gamma^c \lambda.
\end{equation}
Since $\accentset{\neg}{\lambda}\lambda$ is a real scalar, the covariant derivative on the dual spinor $\accentset{\neg}{\lambda}$ can be obtained from the Leibnitz rule. We obtain then
\begin{equation}
\tilde{\nabla}_a \accentset{\neg}{\lambda} = \nabla_a \accentset{\neg}{\lambda} - \frac{1}{4}K_{abc}\accentset{\neg}{\lambda}\gamma^b\gamma^c.
\end{equation}

After defining the covariant derivatives of the {\color{black} dark spinors within a geometry with torsion, we can construct our dark energy model by considering ghost dark spinors; i.e., with a negative kinetic energy,  in an Einstein-Cartan theory, where our lagrangian density reads}
\begin{equation}
\tilde{\mathcal{L}}_{dGS}=-\frac{1}{2}g^{ab}\tilde{\nabla}_{(a}\accentset{\neg}{\lambda} \tilde{\nabla}_{b)}\lambda-V(\accentset{\neg}{\lambda}\lambda),
\end{equation}
where $V(\accentset{\neg}{\lambda}\lambda)$ is an arbitrary potential. Besides, we should mention that the main difference between Ref. \cite{Bohmer2008} and our work is that here we consider a negative kinetic term regarding it as a dynamical dark energy model and we analyze if the model would lead to instabilities or not {\color{black}(on the form of dark energy singularities)}. {\color{black} Notice that if we only use $g^{ab}\tilde{\nabla}_{a}\accentset{\neg}{\lambda} \tilde{\nabla}_{b}\lambda$ in our lagrangian, after taking the variation with respect to the metric, we are left with the term} $\tilde{\nabla}_a \accentset{\neg}{\lambda} \tilde{\nabla}_b \lambda $, which is not necessarily symmetric since the spin connection does not commute with each other in general, i.e. $\Gamma_a\Gamma_b\neq\Gamma_b\Gamma_a$,  {\color{black} even in absence of torsion.} Therefore, we have to symmetrize the kinetic term to ensure the symmetric property of the field equation. Although the lagrangian density is somewhat similar to the one of a complex scalar field, we emphasize that a complex scalar field is a spin-0 field, and hence cannot interact with torsion as a spinor field does. Taking the variation with respect to the metric, we obtain the metric energy-momentum tensor
 \begin{equation}
 \tilde{\sigma}_{ij}= -2 \tilde{\nabla}_{(i}\accentset{\neg}{\lambda}\tilde{\nabla}_{j)}\lambda-g_{ij}\tilde{\mathcal{L}}_{dGS}.
 \end{equation}

The spin tensor can be obtained by taking the variation of the action with respect to the contortion tensor:
 \begin{equation}
 \tau^{kj}{}{}_{i}=\frac{\delta \tilde{\mathcal{L}}_{dGS}}{\delta K^{i}{}{}_{jk}}=-\frac{1}{4}\tilde{\nabla}_i\accentset{\neg}{\lambda} \gamma^j
\gamma^k \lambda + \frac{1}{4}\accentset{\neg}{\lambda} \gamma^j
\gamma^k \tilde{\nabla}_i \lambda,
\end{equation}
which can be separated into torsion-free and non-torsion free parts,
\begin{align*}
\tau^{ij}{}{}_{k}=&-\frac{1}{4} \nabla_k \accentset{\neg}{\lambda}\gamma^j\gamma^i \lambda+\frac{1}{4}\accentset{\neg}{\lambda}\gamma^j\gamma^i \nabla_k \lambda
+\frac{1}{16}K_{kab}\accentset{\neg}{\lambda}\gamma^a \gamma^b \gamma^j\gamma^i \lambda \\
&+\frac{1}{16}K_{kab}\accentset{\neg}{\lambda}\gamma^j \gamma^i \gamma^a\gamma^b \lambda, \tag{3.11}
\end{align*}
where the first two terms on rhs in {\color{black} Eq. (3.11)} are torsion free while the last two terms are non-torsion free. From this, we can see that the spin angular momentum tensor indeed depends on the contortion tensor and cannot be expressed as an axial vector of the torsion tensor as the Dirac spinor does \cite{Hehl1976, Popp2012}. Therefore, the {\color{black} dark ghost} spinor can possibly couple to all the irreducible parts of the torsion tensor, and give richer implications in Einstein-Cartan cosmology than the ordinary Dirac spinors \cite{Bohmer2006}.

From Sec. II, we know that the gravitational action in Einstein-Cartan theory is similar to GR, the difference lies in the Ricci scalar $\tilde{R}$, where we treat the metric and the non-torsion free affine connection to be independent variables. It follows that the full action of our model reads
\setcounter{equation}{11}
\begin{equation}
S=\int d^4x \sqrt{-g}\left(\frac{1}{2\kappa}\tilde{R}+\tilde{\mathcal{L}}_{dGS} \right).
\end{equation}
In a spatially homogeneous and isotropic universe, we use the flat Friedman-Lema\^{i}tre-Robertson-Walker (FLRW) metric
\begin{equation}
ds^2=dt^2-a^2(t)(dx^2+dy^2+dz^2),
\end{equation}
where $a(t)$ is the scale factor. Accordingly, the vielbein $e^\mu_a$ are easy to obtain
\begin{equation}
e^\mu_0 =\delta^\mu_0,\,\,\,e^\mu_i=\frac{1}{a}\delta^\mu_i, \tag{3.14a}
\end{equation}
and the inverse vielbein $e^a_\mu$ reads
\begin{equation}
e^a_0 =\delta^a_0,\,\,\, e^a_i=a\delta^a_0. \tag{3.14b}
\end{equation}
In this background, the non-vanishing torsion free Christoffel symbols are \cite{Bohmer2008}
\begin{align}
&\Gamma^x_{tx}=\Gamma^y_{ty}=\Gamma^z_{tz}=\frac{\dot{a}}{a}, \tag{3.15a}\\
&\Gamma^t_{xx}=\Gamma^t_{yy}=\Gamma^t_{zz}=a\dot{a}, \tag{3.15b}
\end{align}
where the dot denotes differentiation with respect to the cosmic time $t$.
The corresponding spin connection coefficients in the holonomic frame $\Gamma_\mu$ can be obtained by {\color{black} using Eq. (3.6) and read} \cite{Bohmer2008,Bohmer2010}
\setcounter{equation}{15}
\begin{equation}
\Gamma_t=0, \,\,\, \Gamma_{x^i}=-\frac{1}{2}(a\dot{a})\gamma^t\gamma^{x^i},\,x^i=x,y,z.
\end{equation}
It follows that we can compute the spin connection in the anholonomic frame, $\Gamma_a$, {\color{black} and} the non-vanishing terms are
\begin{equation}
\Gamma_0=0, \,\,\, \Gamma_i=-\frac{1}{2}\left( \frac{\dot{a}}{a}\right)\gamma^0\gamma^i,\,i=1,2,3.
\end{equation}
If the cosmological principle is assumed, it can greatly reduce the degrees of freedom of the torsion, in other words, the only not necessarily vanishing components {\color{black} of the} torsion tensor in the anholonomic frame are \cite{Bohmer2008}
\begin{align}
&S_{ijk}=f(t)\epsilon_{ijk}, \tag{3.18a}\\
&S_{i0i}=-h(t),\,\,\, i=1,2,3 \tag{3.18b},
\end{align}
where $f(t)$ and $h(t)$ are called the torsion functions, which depend only on $t$ due to the homogeneous and isotropic {\color{black}assumptions}, and $\epsilon_{ijk}$ is the anti-symmetric Levi-Civita symbol with $\epsilon_{123}=1$.

With the above expression, once we know the non-vanishing torsion terms, we can obtain the non-vanishing contortion terms by means of Eq. (2.2). Then, by using Eq. (2.1) we can determine the connection $\tilde{\Gamma}_{\mu\nu}^\lambda$ and finally compute the Einstein tensor with torsion $\tilde{G}_{ij}$ directly using the definition of the Ricci tensor,
$\tilde{R}_{\sigma\nu}=\partial_\mu \tilde{\Gamma}^\mu_{\nu\sigma}-\partial_\nu\tilde{\Gamma}^\mu_{\mu\sigma}+\tilde{\Gamma}^\mu_{\mu\lambda}\tilde{\Gamma}^\lambda_{\nu\sigma}
-\tilde{\Gamma}^{\mu}_{\nu\lambda}
\tilde{\Gamma}^\lambda_{\mu\sigma}$. {\color{black}Using these steps,} we obtain \cite{Bohmer2008}
\begin{align}
\tilde{G}_{tt}&=3\left(\frac{\dot{a}}{a}\right)^2+12\left(\frac{\dot{a}}{a}\right)h+12h^2-3f^2, \tag{3.19}\\
\tilde{G}_{xx}&=a^2 \left[-2 \left(\frac{\ddot{a}}{a}\right)-\frac{\dot{a}}{a}\left(\frac{\dot{a}}{a}+8h\right)-4\dot{h}-4h^2+f^2 \right], \tag{3.20} \\
\tilde{G}_{xx}&=\tilde{G}_{yy}=\tilde{G}_{zz}. \tag{3.21}
\end{align}

On the other hand, to obtain the complete field equation, one also has to know the energy-momentum tensor, the rhs of Eq. (2.4), $\tilde{\Sigma}_{ij}$. Since the cosmological principle has to be applied not only to the geometrical side but also to the matter side, the matter distribution should be also homogeneous and isotropic. Therefore, we can assume that the {\color{black} dark ghost} spinor fields in our model depend only on time, $t$, writing $\lambda(t)=\varphi(t)\xi$ and $\accentset{\neg}{\lambda}=\varphi(t)\accentset{\neg}{\xi}$, where $\varphi(t)$ is a real function and $\xi$ is a constant {\color{black} dark ghost} spinor and its {\color{black} corresponding}  dual $\accentset{\neg}{\xi}$ is defined by {\color{black} Eq. (3.5)}. Since the cosmological principle implies the off-diagonal components of the Einstein tensor to vanish, for example $\tilde{G}_{tx}=\tilde{G}_{xy}=0$, it naturally constrains the energy-momentum tensor on the rhs of the field equation, Eq. (2.4). That is to say, the off-diagonal components of the energy-momentum tensor should also vanish even in absence of torsion. To be precise, this means that the {\color{black} dark ghost} spinor is required to satisfy the condition that the off-diagonal components of the metric energy-momentum tensor should also vanish, i.e. $\tilde{\sigma}_{tx}=\tilde{\sigma}_{xy}=0.$ {\color{black} The simplest way to satisfy this condition is to assume a spinor with zero norm, $\accentset{\neg}{\lambda}\lambda=0$ \cite{Bohmer2007}}. {\color{black} In this context, the word ``ghost'' refers to the fact that it has no contribution to the metric energy-momentum tensor and thus has no effect on the curvature of spacetime in the absence of torsion \cite{Davis1972,Griffiths2001,Dimakis1982}. In our case the word ``ghost'' can be used, in addition, because of the sign of the kinetic term in the action (3.8)}. A cosmological dark ghost spinor can be given by \cite{Bohmer2008, Bohmer2007}
\setcounter{equation}{21}
\begin{align}
\lambda_{\{-,+\}}     =\varphi(t)\xi, \,\,\, \xi=\left(
                                              \begin{array}{c}
                                                0 \\
                                                \pm i \\
                                                1 \\
                                                0 \\
                                              \end{array}
                                            \right)\\
\lambda_{\{+,-\}}     =\varphi(t)\zeta, \,\,\, \zeta= i \left(
                                              \begin{array}{c}
                                                \mp i \\
                                                0 \\
                                                0 \\
                                                -1 \\
                                              \end{array}
                                            \right)
\end{align}
{\color{black} while its corresponding dual spinor reads}
\begin{equation}
\accentset{\neg}{\lambda}=\varphi(t)\accentset{\neg}{\xi}, \,\,\, \accentset{\neg}{\xi}=i\left(0,i,\pm 1,0 \right).
\end{equation}
\textcolor{black} {Please notice that in Eq. (3.24) and in what follows, we suppress the helicity index, i.e., $\lambda_{\{-,+\}} = \lambda$, because in our model, we assume that $\lambda_{\{-,+\}}$ is the only dark ghost spinor and it corresponds to dark energy in our universe. Therefore, although in principle $\lambda_{\{+,-\}}$ could as well contribute to dark energy, we choose only one spinor of the two possible spinors for simplicity.}
Since the norm of the dark ghost spinor vanishes, the potential $V$ which is a function of $\accentset{\neg}{\lambda}\lambda$ plays a role similar to that of the cosmological constant. Note that Eq. {\color{black}(3.22)} is based on the consistency of the cosmological principle, i.e. our universe is homogeneous and isotropic on large scale, therefore only the time dependence degree of freedom remains. Besides, the energy-momentum tensor has also to be compatible with the cosmological principle, thus the zero-norm is a natural choice\footnote{At cosmological scales, the fermion number might not be conserved, see for example Refs. \cite{Gibbons1979,Jeffrey1990}.}. However, a {\color{black}more general} spinor with non-vanishing norm may also satisfy the condition that the off-diagonal components of the metric energy-momentum tensor vanish, in that case, higher order self-interactions are allowed.

\textcolor{black}{The Cartan equation (2.5) is in general a set of 24 algebraic equations, however, using Eqs. (3.18), it reduces to two independent equations relating torsion and spin tensors, i.e. $T_{123}=f(t)=\kappa\tau_{123},$ and $T_{101}=2h(t)=\kappa\tau_{101}$.
Using Eq. (3.22), we can calculate
\begin{align}
\tau^{12}{}{}_3=&-\frac{1}{4} \nabla_3 \accentset{\neg}{\lambda}\gamma^2\gamma^1 \lambda+\frac{1}{4}\accentset{\neg}{\lambda}\gamma^2\gamma^1 \nabla_3 \lambda \notag
+\frac{1}{16}K_{3ab}\accentset{\neg}{\lambda}\gamma^a \gamma^b \gamma^2\gamma^1 \lambda \\ \notag
&+\frac{1}{16}K_{3ab}\accentset{\neg}{\lambda}\gamma^2 \gamma^1 \gamma^a\gamma^b \lambda,\\ \notag
=&\frac{1}{2}\left(\frac{\dot{a}}{a}\right)\varphi^2+h(t)\varphi^2, \notag\\
\tau^{10}{}{}_1=&-\frac{1}{4} \nabla_1 \accentset{\neg}{\lambda}\gamma^0\gamma^1 \lambda+\frac{1}{4}\accentset{\neg}{\lambda}\gamma^0\gamma^1 \nabla_1 \lambda \notag
+\frac{1}{16}K_{1ab}\accentset{\neg}{\lambda}\gamma^a \gamma^b \gamma^0\gamma^1 \lambda \\ \notag
&+\frac{1}{16}K_{1ab}\accentset{\neg}{\lambda}\gamma^0 \gamma^1 \gamma^a\gamma^b \lambda,\\ \notag
=&\frac{1}{2}f(t)\varphi^2. \notag
\end{align}
Here, we have used the following properties of the ghost dark spinor: $\accentset{\neg}{\lambda} \lambda=0$, $\accentset{\neg}{\lambda}(\gamma^0 \gamma^1 \gamma^2\gamma^3 ) \lambda=\accentset{\neg}{\lambda}(-i\gamma^5) \lambda=-2\varphi^2$, and the non-zero components of the contortion tensor: $K_{123}=-f(t)$ and $K_{101}=K_{202}=K_{303}=2h(t)$.
Then, it is straightforward to solve the functions $f(t)$ and $h(t)$ in terms of the matter field $\varphi(t)$}
\begin{align}
&h(t)=-\frac{1}{4}\kappa \varphi^2 f=-\frac{\frac{1}{2}\kappa^2\varphi^4}{4+\kappa^2\varphi^4}\left(\frac{\dot{a}}{a}\right), \tag{3.25} \\
&f(t)=\frac{2\kappa\varphi^2}{4+\kappa^2\varphi^4}\left(\frac{\dot{a}}{a}\right). \tag{3.26}
\end{align}
Here, we can see that the dark ghost spinor indeed has non-trivial contributions to both the spatial axial components and to the temporal components of the torsion tensor as compared with the Dirac spinor which has only a contribution to the spatial axial vector components of the torsion tensor \cite{Popp2012, Bohmer2006}. Moreover, the non-trivial components of the spin angular momentum tensor in our model are $\tau_{123}=\frac{1}{2}\frac{\dot{a}}{a}\varphi^2+h\varphi^2$ and $\tau_{101}=-\frac{1}{2}f\varphi^2=\tau_{202}=\tau_{303}$, which are of {\color{black}course} homogeneous and isotropic in agreement with the cosmological principle. To obtain the canonical energy-momentum tensor, we need to compute the contributions of the spin angular momentum taking into account the torsion interactions,  $(\tilde{\nabla}_k+2S_{kl}{}{}^l)(\tau_{ij}{}{}^k-\tau_j{}^k{}_i+\tau^k{}{}_{ij})$ (cf. Eqs. (2.2) and (2.7)). Finally, we obtain that the non-vanishing components reads
\begin{align}
\Sigma_{tt}&=V_0+3\left( \frac{\dot{a}}{a}\right)f\varphi^2+6fh\varphi^2, \tag{3.27}\\
\Sigma_{xx}&=-a^2V_0-a^2\varphi^2f\left(6h-2\frac{\dot{\varphi}}{\varphi}-\frac{\dot{f}}{f}\right), \tag{3.28}\\
\Sigma_{xx}&=\Sigma_{yy}=\Sigma_{zz}, \tag{3.29}
\end{align}
where $V_0=V(0)$. We will analyze the dynamics of our dark energy model in the next section.

\section{Cosmological Evolution of the ghost dark Spinor}
The evolution of the Hubble parameter, $H=\dot{a}/a$, can be determined from Einstein equation (2.4). The corresponding Friedmann  and Raychaudhuri equations read
\begin{align}
H&=\frac{\sqrt{\kappa V_0}}{2\sqrt{3}}\frac{4+\kappa^2 \varphi^4}{\sqrt{4-3\kappa^2 \varphi^4}}, \tag{4.1}\\
\dot{H}&=-\frac{\kappa V_0}{12}\frac{20\kappa^2\varphi^4+3\kappa^4\varphi^8}{4-3\kappa^2 \varphi^4}, \tag{4.2}
\end{align}
where the equations (3.19)-(3.21) and {\color{black} (3.27)-(3.29)} have been used.
The evolution of the matter field $\varphi(t)$ can be obtained by taking the time derivative of Eq. (4.1) and equating it to Eq. (4.2), then
\setcounter{equation}{2}
\begin{equation}
\frac{\dot{\varphi}}{\varphi}=-\frac{\sqrt{\kappa V_0}}{4\sqrt{3}}\frac{20+3\kappa^2 \varphi^4}{20-3\kappa^2 \varphi^4}\sqrt{4-3\kappa^2\varphi^4}.
\end{equation}
Eq. (4.3) gives the evolution of the matter field, combining it with Eq. (4.1), one obtains a differential equation for the scale factor in terms of the matter field $\varphi(t)$
\begin{equation}
\frac{d\ln a}{d\ln\varphi}=-2\frac{4+\kappa^2\varphi^4}{4-3\kappa^2\varphi^4}\frac{20-3\kappa^2\varphi^4}{20+3\kappa^2\varphi^4}.
\end{equation}
{\color{black} After solving the above differential equation, we obtain}
\begin{equation}
a(\varphi)= \frac{a_0}{\varphi^2}\left[\frac{(4-3\kappa^2\varphi^4)^4}{20+3\kappa^2\varphi^4}\right]^{\frac{1}{9}},
\end{equation}
where $a_0$ is an integration constant.

{\color{black} As we mentioned previously}, the Einstein-Cartan equation (2.9) can be interpreted as that the geometry is the result from the contribution of the matter fields plus some spin-spin interaction. Therefore for a homogeneous and isotropic Universe, we can define for example an equation of state for dark energy, $w_{d}$, related to the {\color{black}ghost dark} spinor from the metric energy-momentum tensor given in Eq. (2.8) where
$\tilde{\sigma}^i{}_j= \mathrm{diag}(\rho_{d},-p_{d},-p_{d},-p_{d})$, and $w_{d}\equiv p_{d}/\rho_{d}$, then we have
\begin{equation}
w_{d}=-1+\frac{2\kappa^2\varphi^4}{12-3\kappa^2\varphi^4}.
\end{equation}
This equation of state does not take into account the spin-spin interaction; i.e. the energy momentum tensor $u_{ij}$ defined in Eq. (2.10). We could equally define a spin-spin effective equation of state related to $u_{ij}$ which we omit here for simplicity.

Since $\varphi$ is constrained by {\color{black} Eq.(4.3)} to satisfy the condition $0\leq\varphi^2<\sqrt{\frac{4}{3\kappa^2}}$, the time derivative of $\varphi$ is always negative (please c.f. again {\color{black}Eq. (4.3)}), then $\varphi$ will decrease. In fact, $\varphi$ will monotonically decrease to its lower bound, $\varphi =0$, as time goes to infinity, as we next show in Eq. {\color{black} (4.11)}.
\begin{figure}
\includegraphics{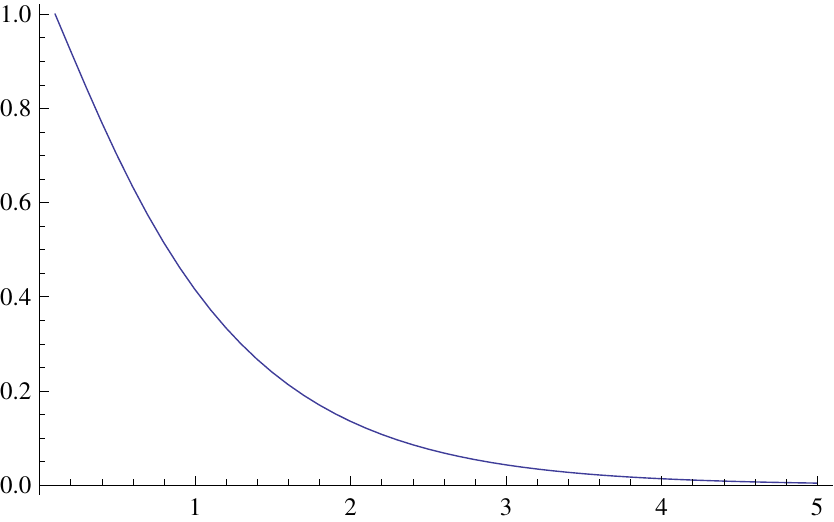}
\caption{Numerical plot of $\varphi(t)$  in Eq. (5.3) from $t=0$ to $t=5$ with $\varphi(0)= 1$, and $\kappa =V_0=1$.}
\end{figure}
Then, $w_{d}$ goes to $-1$ asymptotically, the Hubble parameter is almost a constant, and the scale factor expands as $a(t) \varpropto \exp(Ht)$, therefore our universe enters a de Sitter phase at late time. From the positivity of the second term on rhs in {\color{black}  Eq. (4.6)}, we see that the equation of state will be always larger than -1. And if we consider the contribution of spin-spin interaction, we can define the total equation of state $w_{tot}\equiv p_{tot}/\rho_{tot}$ from $\hat{\sigma}^{i}{}_{j}=\mathrm{diag}(\rho_{tot},-p_{tot},-p_{tot},-p_{tot})$, then
\begin{equation}
w_{tot}=-1+\frac{2}{3}\,\frac{20\kappa^2\varphi^4+3\kappa^4\varphi^8}{(4+\kappa^2\varphi^4)^2}.
\end{equation}
Since both definitions {\color{black} for the} equation of state show that our dynamical dark energy model does not cross the phantom divide, we do not expect quantum instabilities even though the kinetic energy is negative, {\color{black} cf. Eq. (3.8)}.
Note that in a finite cosmic time, $\varphi$ will never become zero, therefore,  neither the Hubble parameter nor its time derivative diverge at a finite cosmic time, hence this model is free from the big rip singularity \cite{Starobinsky2000,Caldwell2003,Caldwell2002,Gonzalez-Diaz2004,Gonzalez-Diaz2004a}. Indeed, the universe would be asymptotically de Sitter in this model.

We can expand {\color{black} Eqs. (4.1) and (4.3)} around $\varphi=0$ to the first few orders to see its qualitative behavior,
\setcounter{equation}{7}
\begin{gather}
H=\frac{\sqrt{\kappa V_0}}{2\sqrt{3}}\left(1+\frac{5}{4}\kappa^2\varphi^4+\mathcal{O}(\varphi^8)\right),\\
\frac{\dot{\varphi}}{\varphi}=-\frac{\sqrt{\kappa V_0}}{2\sqrt{3}}\left(1-\frac{3}{40}\kappa^2\varphi^4+\mathcal{O}(\varphi^8)\right).
\end{gather}
Solving these differential equations to first order, we obtain
\begin{align}
a(t)&=a_0\exp{\left(\frac{\sqrt{\kappa V_0}}{2\sqrt{3}}t\right)}, \tag{4.10}\\
\varphi(t)&=\varphi_0\exp{\left(-\frac{\sqrt{\kappa V_0}}{2\sqrt{3}}t\right)}. \tag{4.11}
\end{align}
We see that as time goes to infinity, $\varphi(t)$ exponentially decays, so do the torsion functions $h$ and $f$; {\color{black} cf. Eqs. (3.24) and (3.25)}. It is not surprising because as the spin sources dilutes the torsion will vanish accordingly \cite{Bohmer2008}.

We next consider the existence of some kind of cold dark matter given by a perfect fluid of a spin-0 particle with the energy-momentum tensor given by ${\sigma_{(m)}}^{\mu}{}_{\nu}=\mathrm{diag}(\rho_m,0,0,0)$, where $\rho_m$ is its energy density. Since it has spin zero, it has no extra contribution to the torsion by the Cartan equation, Eq.(2.5), it only has an additive contribution to the total energy-momentum tensor, $\hat{\sigma}_{ij}$ in Eq. (2.9), that is
\setcounter{equation}{11}
\begin{equation}
\hat{\sigma}_{ij}=\sigma^{(m)}_{ij}+\tilde{\sigma}^{(de)}_{ij}+\kappa u_{ij},
\end{equation}
where $\tilde{\sigma}^{(de)}$ is the metric energy-momentum tensor of the {\color{black} ghost dark} spinor defined in Eq. (2.8).
Then {\color{black} Eqs. (4.1) and (4.3)}, are modified as
\setcounter{equation}{11}
\begin{align}
H^2& =\frac{\kappa V_0}{12}(1+\beta)\frac{(4+\kappa^2\varphi^4)^2}{4-3\kappa^2\varphi^4}, \\
\frac{\dot{\varphi}}{\varphi}& =-\frac{\sqrt{\kappa V_0}}{4\sqrt{3}}\frac{1}{\sqrt{1+\beta}}\frac{20+3\kappa^2 \varphi^4}{20-3\kappa^2 \varphi^4}\sqrt{4-3\kappa^2\varphi^4} \notag\\
&-\frac{1}{4}\frac{(4+\kappa^2\varphi^4)(4-3\kappa^2\varphi^4)}{20\kappa^2\varphi^4-3\kappa^4\varphi^8}\frac{\dot{\beta}}{1+\beta},
\end{align}
where $\beta\equiv \frac{\rho_m}{V_0}$ while {\color{black} Eq. (4.2)} remain unchanged.
We define the total equation of state of the universe by using again $\hat{\sigma}^i{}_{j}=\mathrm{diag}(\rho_{tot},-p_{tot},-p_{tot},-p_{tot})$, which gives
\begin{equation}
w_{tot}\equiv \frac{p_{tot}}{\rho_{tot}}=-1+\frac{2}{3}\,\frac{20\kappa^2\varphi^4+3\kappa^4\varphi^8}{(4+\kappa^2\varphi^4)^2}(1+\beta)^{-1}.
\end{equation}

The conservation of the energy-momentum tensor $\nabla_i\sigma^{(m)i}{}_{j}=0$ reads
\setcounter{equation}{14}
\begin{equation}
\frac{\dot{\beta}}{\beta}=-3\left(\frac{\dot{a}}{a}\right).
\end{equation}
Substituting $\frac{\dot{a}}{a}$ using {\color{black} Eq. (4.12) into Eq. (4.15)}, we get
\begin{equation}
\dot{\beta}=-\frac{\sqrt{3\kappa V_0}}{2}\frac{4+\kappa^2\varphi^4}{\sqrt{4-3\kappa^2\varphi^4}}\beta\sqrt{\beta +1}.
\end{equation}

To see the stability of the late time behavior, we analyze the autonomous $(\varphi,\beta)$ system , which is
\begin{align}
\dot{\varphi}&=-\frac{\sqrt{\kappa V_0}}{4\sqrt{3}}\frac{\varphi\sqrt{4-3\kappa^2\varphi^4}}{\sqrt{1+\beta}}\frac{20+3\kappa^2\varphi^4}{20-3\kappa^2\varphi^4} \notag\\
&+\frac{\sqrt{3\kappa V_0}}{8}\frac{\varphi(4+\kappa^2\varphi^4)^2\sqrt{4-3\kappa^2\varphi^4}}{20\kappa^2\varphi^4-3\kappa^4\varphi^8}\frac{\beta}{\sqrt{1+\beta}},\\
\dot{\beta}&=-\frac{\sqrt{3\kappa V_0}}{2}\frac{4+\kappa^2\varphi^4}{\sqrt{4-3\kappa^2\varphi^4}}\beta\sqrt{\beta +1}.
\end{align}
The only fixed point is $(\varphi_0,\beta_0)=(0,0)$. We linearize the system around the fixed point, by expanding $(\varphi,\beta)=(\varphi_0+\delta\varphi,\beta_0+\delta\beta)$, and we obtain that
\begin{equation}
\binom{\delta\dot{\varphi}}{\delta\dot{\beta}}=\frac{\sqrt{\kappa V_0}}{2\sqrt{3}}\left(
                                                 \begin{array}{cc}
                                                   -1 & 0 \\
                                                   0 & -6 \\
                                                 \end{array}
                                               \right)\binom{\delta\varphi}{\delta\beta}.
\end{equation}
The linearized system is automatically diagonal, one can easily read off its eigenvalues, both are real and negative.
Therefore, $(\varphi,\beta)=(0,0)$ is an attractive fixed point, and this would give us $w_{tot}\rightarrow-1$ in the future. As the universe expands, the torsion will vanish. When both $\varphi$ and $\beta$ are small, the Hubble parameter will be nearly constant, the scale factor $a(t)$ grows exponentially which means the universe will again enter a de Sitter phase.

The numerical evolution of the equation of state, $w_{tot}(z)$, and the Hubble parameter $H(z)$ of the universe with redshift $z \equiv -1+\frac{a_0}{a}$ {\color{black} are shown in Figs. 2 and 3, respectively,} where $a_0$ stands for the present value of the scale factor.

Note that for $\tilde{\sigma}^{(de)i}{}_j=\mathrm{diag}(\rho_{de},-p_{de},-p_{de},-p_{de})$, the conservation equation, $\nabla_i(\tilde{\sigma}^{(de)i}{}_j+\kappa u^i{}_j)=0$, can be interpreted as the continuity equation of the energy density of the ELKO spinor with a source term, $\dot{\rho}_{de}+3H(\rho_{de}+p_{de})=Q$ where $Q>0$ means energy is transferred from torsion to {\color{black} ghost dark spin}, and $Q=0$ means no interaction between torsion and {\color{black} the spin field}.

We can as well define an equation of state for dark energy again as $w_{de}\equiv \frac{p_{de}}{\rho_{de}}$, then\footnote{{\color{black}Please note that Eq. (4.20) is different from Eq. (4.6). The reason is that by adding cold dark matter into the model, we modify the spin connection (cf. Eq. (3.16) and (3.17)), therefore we modify the spin energy momentum tensor in Eq. (3.9).}}
\begin{equation}
w_{de}=-1+\frac{2\kappa^2\varphi^4(1+\beta)}{3(4-3\kappa^2\varphi^4)+6\kappa^2\varphi^4(1+\beta)}.
\end{equation}
We can as well define an effective equation of state for dark energy $w^{eff}_{de}\equiv \frac{p_{de}}{\rho_{de}}-\frac{Q}{3H\rho_{de}}$, then
\begin{align*}
w^{eff}_{de}&=-1+\frac{2\beta\kappa^2\varphi^4}{4-(1-2\beta)\kappa^2\varphi^4} \\ \notag
&+\frac{8}{3}\frac{(6-3\beta)\kappa^4\varphi^8+(40-24\beta)\kappa^2\varphi^4+48}
{(4+\kappa^2\varphi^4)(4-(1-2\beta)\kappa^2\varphi^4)(20-3\kappa^2\varphi^4)}. \tag{4.21}
\end{align*}
The term, $Q$, is equally present when $\beta=0$; i.e. in the absence of dark matter.
The numerical evolution of $w_{de}$ and $w^{eff}_{de}$ with redshift $z$ are given in Figs. 4, and 5. Note that at early time, $w^{eff}_{de}<w_{de}$ which means $Q>0$, thus energy is transferred from torsion to the ELKO fields, and at late time, $w^{eff}_{de}\approx w_{de}$ which means $Q\approx 0$ as is expected since torsion will eventually vanish.

\begin{figure}
\includegraphics{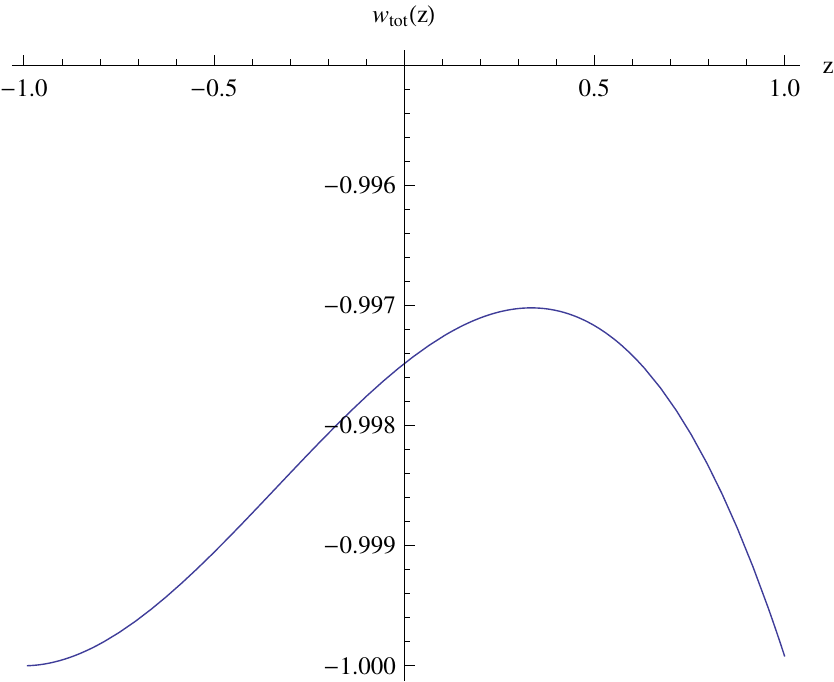}
\caption{$w_{\mathrm{tot}}(z)$ defined in Eq. (5.14) from $z=1$ to $z=-1$ with $\varphi(1)= 0.1$, $\beta(1)=0.01$, $\kappa=1$, and $\Omega_{m_0}\approx 0.3$.}
\end{figure}

\begin{figure}
\includegraphics{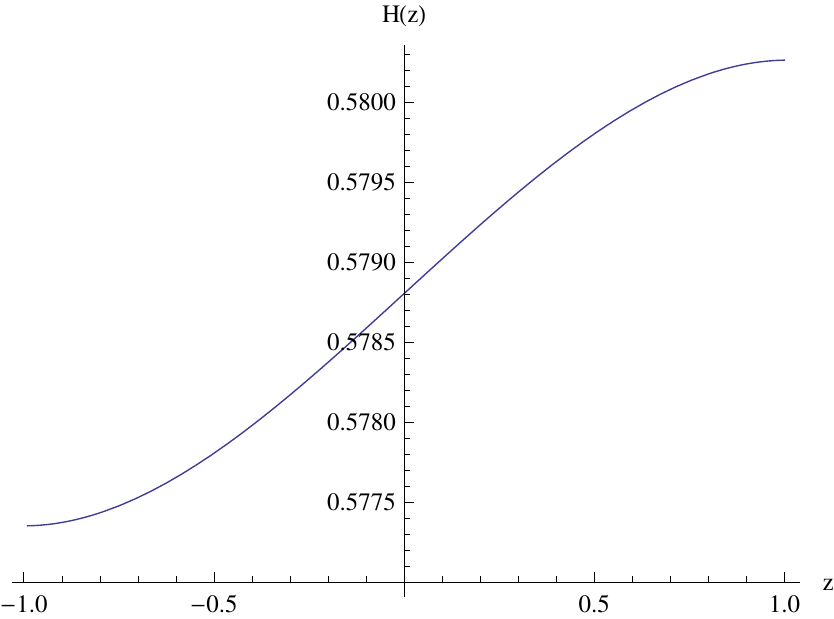}
\caption{$H(z)$ gievn in Eq. (5.12) from $z=1$ to $z=-1$ with $\varphi(1)= 0.1$, $\beta(1)=0.01$, and $\kappa=1$. The asymptotic line is $H(z)=\frac{\sqrt{3}}{3}\approx 0.577$, and $\Omega_{m_0}\approx 0.3$.}
\end{figure}

\begin{figure}
\includegraphics{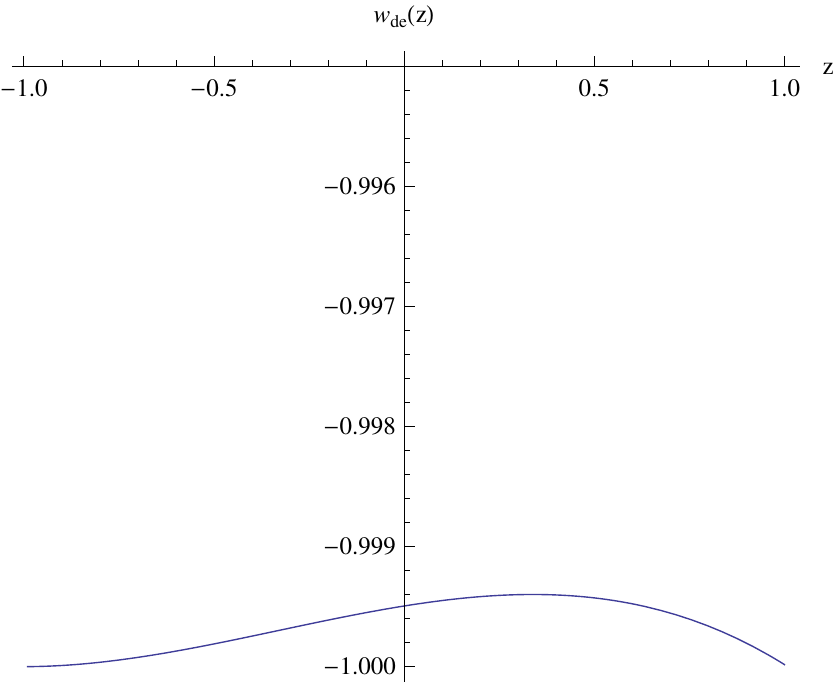}
\caption{$w_{de}(z)$ given in Eq. (5.20) from $z=1$ to $z=-1$ with $\varphi(1)= 0.1$, $\beta(1)=0.01$, $\kappa=1$, and $\Omega_{m_0}\approx 0.3$.}
\end{figure}

\begin{figure}
\includegraphics{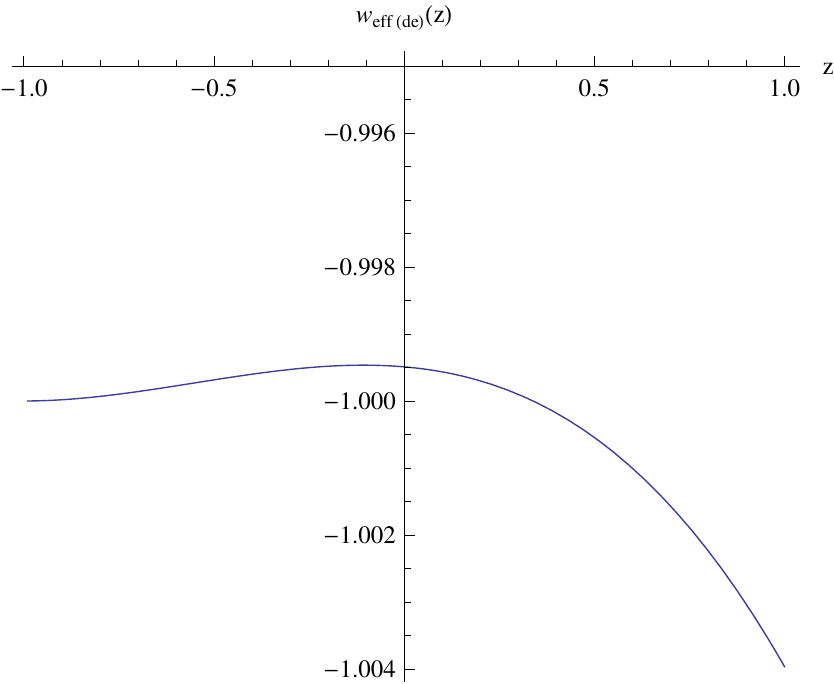}
\caption{$w^{eff}_{de}(z)$ given in Eq. (5.21) from $z=1$ to $z=-1$ with $\varphi(1)= 0.1$, $\beta(1)=0.01$, $\kappa=1$, and $\Omega_{m_0}\approx 0.3$.}
\end{figure}

\section{Conclusion and Discussion}
{\color{black} In this paper, we consider Einstein-Cartan theory, which is the simplest generalization of ordinary general relativity that incorporate  torsion fields as the anti-symmetric part of the affine connection}. In such a theory, there are two field equations, one is like the traditional Einstein equation and the other {\color{black} one is} an algebraic relation between the torsion fields and the spin fields of the matter sources. We introduce a new kind of spin one-half particle called dark spinor which is the eigenspinor of the charge conjugation operator, and different from the Majorana spinor due to the double-helicity structure \cite{Ahluwalia-Khalilova2005}. The equation of motion {\color{black} for such a}  spinor is the Klein-Gordon equation rather than the Dirac equation. Then, we propose a dark energy model with a negative kinetic energy constructed from {\color{black} such a dark spinor which interacts with the torsion fields in a FLRW universe.}

Although the kinetic energy is negative, the equation of states $w_{de}$ and $w_{tot}$ do not cross the phantom divide and approaches $-1$ asymptotically, satisfying the weak energy condition, hence we expect the model to be stable at the quantum level. No big rip singularity will occur at a finite cosmic time in this setup. Torsion will vanish at late time, and the Hubble parameter will approach a constant asymptotically. Furthermore, we consider the existence of some cold dark matter which is assumed to be a pressureless scalar particle without contribution to the torsion fields. In this two components system, we find that there is a unique attractive fixed point, which is simply $(\varphi,\beta)=(0,0)$, and all of the equations of state $w_{tot}$, $w_{de}$, and $w^{eff}_{de}$ will converge to $-1$ from above no matter what the initial condition is. Therefore, the universe will eventually enter a de Sitter phase at late time with or without dark matter.

On this work, we assumed a constant potential in Eq. (4.8) as it is the simplest way to fulfil the requirement of homogeneity and isotropy. Therefore, we did not consider the spinor mass, even though evolving potentials can be considered and we will leave it for a next work \cite{Chang2015}. This model could equally help to solve the coincidence problem as in principle dark matter would have spin which would interact with torsion, which itself would interact with the spinor which plays the role of dark energy. Here also we leave this issue as a future work \cite{Chang2015a}.

{\color{black} The main difference between our work and reference \cite{Bohmer2008} is precisely the sign of the kinetic term of the spinor. In addition, the authors of Ref. \cite{Bohmer2008} identified a matter coupling source whose spin-angular momentum is compatible with a homogeneous and isotropic space-time and in particular they found very interesting inflationary solutions of de Sitter type, i.e. they applied their model to physics of the early universe unlike us that were more interested on the late-time cosmology and on potential ways of removing dark energy singularities. In addition, as shown in Fig. 5, our theory exhibits ``safe'' phantom-like behaviours; i.e. we obtain a phantom-like behaviour in the absence of a dark energy singularity.}

\begin{acknowledgments}
The work of MBL is supported by the Basque Foundation of Science Ikerbasque. She wishes to acknowledge the partial support from the Basque government Grant No. IT956-16 (Spain) and FONDOS FEDER under grant FIS2014-57956-P (Spanish government). She also wishes to acknowledge the hospitality of LeCosPA Center at the National Taiwan University during the completion of part of this work. Y.-C. C. and P.C. are supported by Taiwan National Science Council under Project No. NSC 97-2112-M-002-026-MY3 and by Taiwans National Center for Theoretical Sciences (NCTS). P.C. is in addition supported by US
Department of Energy under Contract No. DE-AC03-76SF00515.
\end{acknowledgments}

\bibliographystyle{unsrt}

\end{document}